\def\ba{\begin{eqnarray}}
\def\ea{\end{eqnarray}}
\def\lb{\label}

\documentclass[12pt]{article}
\usepackage{graphicx}
\begin{document}
\title{Finite size of hadrons and Bose-Einstein correlations  }

\author{A.Bialas$^{a,b}$ and K.Zalewski$^{a,b}$
\\$^{a}$ H.Niewodniczanski Institute of Nuclear Physics\\Polish Academy of Sciences\thanks{Address: Radzikowskiego 152, 31-342 Krakow, Poland}
\\$^{b}$ M.Smoluchowski Institute of Physics\\Jagellonian University\thanks{Address: Reymonta 4, 30-059 Krakow, Poland; e-mails: bialas@th.if.uj.edu.pl, zalewski@th.if.uj.edu.pl}
} \maketitle

\begin{abstract}

It is observed that the finite size of hadrons produced in high energy
collisions implies that their positions are correlated, since the probability to find
two hadrons on top of each other is highly reduced. It is then shown that this
effect can naturally explain  the values of the correlation function below one,
observed at LEP and LHC for pairs of identical pions.

\end{abstract}

{\bf 1.} Momentum correlations between identical bosons have been studied for over 50
years \cite{gg,rev,lisa}. In particular for pairs of
identical bosons one usually discusses the correlation function

\begin{equation}\label{}
    C(\textbf{p}_1,\textbf{p}_2) = \frac{{dN}/{d^3p_1d^3p_2}}{{dN}/{d^3p_1}{dN}/{d^3p_2}}.
\end{equation}
The various corrections which have to be applied when extracting this function
from the data are described in the reviews \cite{rev,lisa}. After these
corrections have been applied, it is hoped (and usually taken for granted) that the correlation function contains no other correlations than those  due to Bose-Einstein statistics. In the following section we recall that this implies the inequality

\begin{equation}\label{}
    C(\textbf{p}_1,\textbf{p}_2) \geq 1.
\end{equation}

Actually the data from LEP \cite{lep,lep2} and also recent measurements at LHC
\cite{cms,alice} show consistently a broad minimum where the correlation
function takes values below one
This means that the produced particles are correlated. It is of course
interesting to look for possible origins of these correlations.

The suggestion that such a minimum results from non-resonant, strong final state $\pi^\pm-\pi^\pm$ interactions,
was made by Bowler \cite{bowler}. To estimate  the effect he used as input the measured S-wave, I=2 $\pi-\pi$
phase shift. Later he pointed out, however, that presence of many particles  tends to cancel the effects of
final-state interactions \cite{bowler2}.

More recently, the importance of the observation of such a minimum in
$e^+-e^-$ data was  realized and studied  in \cite{cs}, using the $\tau$-model \cite{tt}.
This was continued by W.J.Metzger in several contributions \cite{metz}.

In this  note we observe that there is a natural source of inter-hadron
correlations following from the fact that non-interacting hadrons, being
composite, extended  objects, cannot be located in space too close to each
other\footnote{In statistical models this is often taken into account as  the
so-called  excluded volume effect, first considered in \cite{hr} and then
studied by many authors \cite{yen,gaz}.}. The point is that, when the two pions
are too close to each other, their constituents mix and they  cannot be
considered as pions subjected to Bose-Einstein statistics. Note that this fact
is not related either to strong interactions or to position-momentum
correlations. This observation obviously implies that the hadron positions, as
measured by quantum interference, are correlated. Using a simple model we
discuss consequences of this kind of correlations on the measurements of the
quantum interference and show that they naturally lead to values below one for
the correlation function, as observed in data \cite{lep,lep2,cms,alice}.

In the following two sections we introduce the notation and give some basic
formulae. In Section 4 we propose a simple model implementing the requirement
that two hadrons cannot be too close to each other.  A summary and some
comments are given in the last section.

{\bf 2.} Let us denote by $\rho$ the single-particle density matrix in the momentum
representation. Then the single particle momentum distribution is

\begin{equation}\label{}
    P(\textbf{p};t) = E\frac{dN}{d^3p}(t) = \rho(\textbf{p},\textbf{p};t).
\end{equation}
Note that the density matrix is here normalized to the total number of
particles $N$ and not to unity. As seen from this formula, the diagonal
elements of the density matrix are non-negative. For uncorrelated particles the
two-particle density matrix is a product of the single-particle ones. For two
identical bosons, however, this product has to be symmetrized and one finds

\begin{equation}\label{}
    P(\textbf{p}_1,\textbf{p}_2;t) = \rho(\textbf{p}_1,\textbf{p}_1;t)\rho(\textbf{p}_2,\textbf{p}_2;t) +
    \rho(\textbf{p}_1,\textbf{p}_2;t)\rho(\textbf{p}_2,\textbf{p}_1;t).
\end{equation}
This yields

\begin{equation}\label{corfun}
    C(\textbf{p}_1,\textbf{p}_2;t) = 1 +
    \frac{|\rho(\textbf{p}_1,\textbf{p}_2;t)|^2}{\rho(\textbf{p}_1,\textbf{p}_1;t)\rho(\textbf{p}_2,\textbf{p}_2;t)}
    \geq 1,
\end{equation}
where we have used the hermiticity of the density matrix which implies that
$\rho(p_2,p_1;t) = \rho^*(p_1,p_2;t)$.

{\bf 3.} In order to discuss the geometry of the interaction region it is necessary to
express the density matrix by an emission function which can be interpreted as
the distribution of particles  in momentum {\it and} position.  For the
single-particle density matrix, the standard choice \cite{lisa} is

\begin{equation}\label{emifun}
    \rho(p,p') = \int d^4x S(x,P)e^{iqx},
\end{equation}
where   $ P = \frac{1}{2}(p+p'),\qquad q =p-p'$.

Similarly, for the two-particle density matrix one has \cite{bk}
\ba
\rho(p_1,p_2;p_1',p_2')=\int d^4x_1d^4x_2 S(P_1,P_2; x_1,x_2)e^{i
q_1x_1+q_2x_2}\lb{ro2}
\ea
Note that the popular interpretation of emission
functions as probability distributions makes sense only if all the momenta
satisfy the on-shell condition $p^2 = m^2$. The assumption that the
four-momentum $P$ can be modified to satisfy this condition without
significantly distorting the results is known as the smoothness assumption
\cite{lisa}.

Using these formulae one obtains for the correlation function
\begin{equation}\label{lis}
    C(p_1,p_2) = \frac{\int d^4x_1\int d^4x_2S(p_1,p_2;x_1,x_2) +\int d^4x_1\int d^4x_2\cos[q(x_1-x_2)]S(P,P;x_1,x_2)}
     {\int d^4x_1 S(p_1,x_1)\int d^4x_2S(p_2,x_2)}.
\end{equation}
If particles are uncorrelated, $S(p_1,p_2;x_1,x_2)=S(p_1,x_1)S(p_2,x_2)$ and
the first term on the right-hand side equals 1. As shown above, in this case
the second term is non-negative. In order to explain the motivation of the
present work let us make the admittedly unrealistic assumption that the
particles are correlated so that $x_1-x_2$ is constant. Then the cosine can be
taken outside the integral and the second terms gets a cosinusoidal dependence
on $q$, taking both positive and negative values. The excluded volume approach
is a somewhat more realistic way of realizing a qualitatively similar scenario.

{\bf 4.} As already mentioned, taking into account the finite size of
hadrons implies that the positions of hadrons in space are correlated. In this
section we show, using a very simple model,  how  these correlations may lead
to a minimum (below 1) in the correlation function of identical particles.

We assume that the correlations affect neither the single particle momentum
distribution nor the unsymmetrized two-particle momentum
distribution\footnote{This condition is satisfied, in particular, when
there is no correlation between the momenta and positions of the produced particles. In elementary collisions this seems a reasonable approximation, although it may be questioned for heavy ion collisions.}.
Therefore  we have

\begin{equation}\label{norcor}
\int d^4x_1d^4x_2S(p_1,x_1,p_2,x_2) = \int d^4x_1S(p_1,x_1)\int d^4x_2S(p_2,x_2).
\end{equation}
As our illustrative example we choose

\begin{equation}\label{the}
    S(p_1,x_1;p_2,x_2) = {\cal N}S(p_1,x_1)S(p_2,x_2)
    \Theta[\textbf{x}_-^2 + t_-^2 - r_0^2],
\end{equation}
where $x_- = x_1-x_2$. The normalizing constant ${\cal N}$ is necessary to
enforce condition (\ref{norcor}). For $t_1=t_2$ the $\theta$-function just
excludes the configurations with $|\textbf{x}_-| < r_0$. The time difference in
the argument corrects for the fact that this exclusion is not needed when one
particle is far in time from the other.

For our illustrative calculation we choose

\begin{equation}\label{}
S(p,x)=e^{-\textbf{x}^2/R^2}e^{-t^2/\tau^2}f(p),
\end{equation}
where the momentum dependent factor is left unspecified. Including the
correlations as in (\ref{the}), substituting into (\ref{lis}) and performing
the Gaussian integrations over $x_+=(x_1+x_2)/2$ in the numerator and over
$x_1$ and $x_2$ in the denominator, one is left with the four-dimensional
integral over $x_-$. Rewriting the integral $d^3x_-$ in spherical coordinates
the integrations over angles can be done. Thus there are only two integrations
left and one gets

\begin{equation}\label{corela}
C(p_1,p_2) - 1 \sim \int_0^\infty dt\int_0^\infty dr
r^2e^{-t^2/2\tau^2}e^{-r^2/2R^2}\frac{\sin Qr}{Qr}\cos(q_0t)\theta(r^2+t^2-r_0^2),
\end{equation}
where $r = |\textbf{x}_-|$, $t=|t_-|$  and $Q = |\textbf{q}|$. The proportionality
coefficient, skipped on the right-hand side, is positive and the normalization is fixed by (\ref{norcor}) which implies $C(p_1=p_2)=2$. The integral in (\ref{corela}) can be easily evaluated numerically.

Since  $2qP = m_1^2 - m_2^2 = 0$ (the on shell condition) one has $q_0 = \textbf{qP}/P_0$ .
Therefore the result depends on the direction of the vector $\bf{q}$. For the "side"
direction $\textbf{qP}=0$ For the "out" direction $\textbf{qP}/P_0= |v| Q\approx Q$
where the approximate equality holds when the particle pair is highly
relativistic.

\begin{figure}[h]
\begin{center}
\includegraphics[scale=1.0]{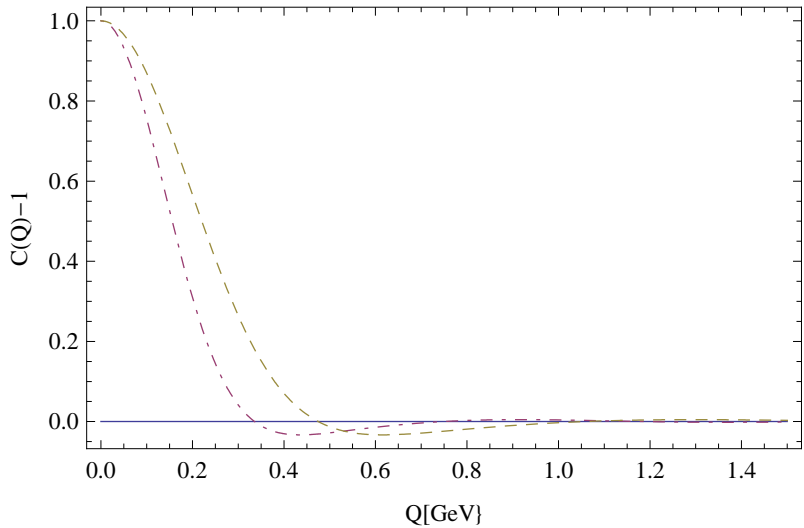}
\end{center}
\caption{ $C(p_1,p_2)-1$ plotted vs $Q$ in GeV. Dashed line: "side" direction; Dashed-dotted line: "out" direction. $r_0=R=\tau=1$ fm.}
\label{th}%
\end{figure}

As an example, Fig. \ref{th} shows the (normalized to 1 at $Q=0$) integrals from  (\ref{corela})
corresponding to the out and side directions,
evaluated at $r_0=R=\tau=1$fm. In both cases one clearly sees a range where the
values of $C(p_1,p_2)-1$ are negative.

Without pretending that this description is realistic, we can thus conclude
that the result proves the main point of our paper: the values below one of the
correlation function describing the quantum interference of  two identical
pions, find a  natural explanation as a consequence of the "excluded volume
effect", i.e. finite size of the produced hadrons.

{\bf 5.} To conclude: starting from the observation that non-interacting hadrons, being  extended objects,  cannot be located  too close to each other, we have investigated the
implications of this fact for the quantum interference of identical bosons.
This "excluded volume" effect implies obviously specific correlations between
particle positions, which are not included in the standard analyses. We have
found that these correlations can lead to the existence of a region where the
observed two-particle correlation function falls below one. This gives, in our
opinion, a natural explanation of the observations reported in several
experiments in $e^+-e^-$ and $p-p$ collisions \cite{lep,lep2,cms}.

 Our treatment of the problem is, admittedly, rather crude, as it involves several approximations
 which are used to avoid complications and thus to  focus  on the two  main points of this paper, that is

 (i) to emphasize the role of inter-hadron correlations in the explanation of the observed negative values of $C(p_1,p_2)-1$ and

 (ii) to point out that a natural source of such inter-hadron correlations can be provided by the finite sizes of the produced hadrons.

Several comments are in order.

(i) Our use of the $\Theta$-function to parametrize the excluded volume
correlations is clearly only a crude approximation. For a precise description
of  data almost certainly a more sophisticated parametrization of the effect
will be needed. In particular, note that with our parametrization the
correlation in space-time does not affect the single particle and two-particle
non-symmetrized momentum distributions. The same comment applies to our use of
Gaussians.

(ii) It has been recently  found \cite{cms,alice} that in $pp$ collisions at
LHC, the volume of the system (as determined from the fitted HBT parameters) depends  weakly on the multiplicity of the particles produced in the collision. This suggests that large multiplicity in an event is due to a longer emission time. If true, this should be also
reflected in the HBT measurements and it may be interesting to investigate this
aspect of the problem in more detail.

(iii) To investigate further the space and/or time correlations between the
emitted particles more information is needed. It would be interesting to study
the minima in the correlation functions separately for the "side", "out" and
"long" directions. Such studies may allow to determine the size of the "excluded volume" and compare it with other estimates \cite{yen,gaz}. We also feel that with the present accuracy and statistics of data, measurements of three-particle B-E correlations represent the potential to provide some essential information helping to understand what is
really going on.

\section{Acknowledgements} Thanks are due to Wojciech Florkowski  and to Krzysztof Redlich for a helpful correspondence.

\end{document}